\documentclass[useAMS,usenatbib]{mn2e}
\usepackage{amssymb,graphics,epsfig,subfigure}

\newcommand{\ciii}{\textsc{C\,iii]}}
\newcommand{\mgii}{Mg\,\textsc{ii}}

\title[The size of the BELR in 2237+0305]{A microlensing measurement of the size of the broad emission line region in the lensed QSO 2237+0305}
\author[R. B. Wayth, M. O'Dowd and R. L. Webster]
{R. B. Wayth$^{1}$\thanks{E-mail: rwayth,modowd,rwebster@physics.unimelb.edu.au },
M. O'Dowd$^{1}$
and R. L. Webster$^{1}$\\
$^{1}$ School of Physics, University of Melbourne, 3010. Australia}
\begin{document}

\date{Accepted -. Received --; in original form --}

\pagerange{\pageref{firstpage}--\pageref{lastpage}} \pubyear{2004}

\maketitle

\label{firstpage}
\begin{abstract}
We present spatially resolved spectroscopic images of the gravitationally lensed QSO 2237+0305 taken with the GMOS Integral Field Unit (IFU) on the Gemini North telescope. These observations have the best spatial resolution of any IFU observations of this object to date and include the redshifted \ciii\ and \mgii\ QSO broad lines.
Unlike \citet{1998ApJ...503L..27M}, we find no evidence for an arc of resolved broad line emission in either the \ciii\ or \mgii\ lines.

We calculate the image flux ratios of both the integrated emission lines and the surrounding continua.
The flux ratios of the \ciii\ and \mgii\ emission lines are consistent with each other but differ substantially from their corresponding continuum flux ratios and the radio/mid-IR flux ratios previously published.
We argue that the broad emission line region must be microlensed and the \ciii\ and \mgii\ emission regions must be approximately the same size and co-located along the line-of-sight.
Assuming a simple model for the broad emission line region and the continuum region, we show the size of the \ciii /\mgii\ broad line region is
$\sim 0.06 h_{70}^{1/2}$ pc and the continuum region is $\le 0.02 h_{70}^{1/2}$ pc.
\end{abstract}


\section{Introduction}\label{sec:intro}

Determining the structural properties of quasars (QSOs) remains a long standing problem in astrophysics.
Reverberation mapping techniques \citep{1997.book.peterson} have proven effective in determining the size of the broad emission line region (BELR) in low-luminosity QSOs.
However, the uncertainties associated with reverberation mapping \citep[see discussion in Sec. 4.1 of][]{2000ApJ...533..631K}, make an independent method for determining the size of QSO emission regions desirable.
Gravitational microlensing offers the opportunity to study the object being lensed (the `source', i.e the QSO) in a unique way.
Using microlensing, it has been possible to deduce information about QSO emitting regions from the light-curves of the lensed images.

\citet{1988ApJ...335..593N} and \citet{1990A&A...237...42S} considered the effect on QSO BELRs by single and multiple microlenses, finding the change in the magnitude and shape of the lines would be relatively small with a large BELR predicted by early photoionisation models ($\sim 0.3$ pc).
Early reverberation mapping results \citep[e.g.][]{1985ApJ...298..283P} and subsequent improvement in QSO models \citep{1989ApJ...347..640R} revised the size of the BELR downwards.
More recent reverberation mapping (including high-luminosity AGN), \citep{2000ApJ...533..631K} suggest the (BELR) is stratified and has a size ranging from a few light-days for Seyfert galaxies to hundreds of light-days for QSOs.
Evidence for a smaller BELR has prompted the idea of BELR microlensing to be revisited.
In general, the magnitude of differential magnification of the BELR and continuum region depends on both the optical depth in compact lenses and the size of the projected microlens Einstein radius relative to the source \citep{1987PhRvL..59.2814D}.
\citet{2002ApJ...576..640A} showed that microlensing of the BELR can be important in the case of a single microlens object.
\citet{2004MNRAS.348...24L} studied the effects in the high optical depth regime and showed the geometry of the BELR can also contribute to the magnifications.


Q2237+0305 \citep[the `Einstein Cross', discovered by ][]{1985AJ.....90..691H} is a $z=1.70$ QSO gravitationally lensed into four images of separation $\sim$1 arcsec around the centre of a $z=0.04$ Sab galaxy.
\citet{1989AJ.....98.1989I} showed that the stars in the bulge of the lens galaxy are microlensing the QSO, a phenomenon which is enhanced in this system by the low redshift of the lens galaxy.
Hence, this object has been the source of numerous studies
to measure properties of the QSO continuum source \citep{1990ApJ...358L..33W,1992ApJ...395L..65R,1999PASJ...51..497M,2000MNRAS.315...62W,2002ApJ...579..127S,2003A&A...397..517G}.

\citet[][hereafter ``M98'']{1998ApJ...503L..27M} found an arc of extended \ciii\ ($\lambda = 190.9$ nm) emission in Q2237+0305 using the `INTEGRAL' integral field unit (IFU) on the William Herschel Telescope.
Using a simple lens model, they showed that existence of this arc implies the \ciii\ BELR in Q2237+0305 is $\sim1$kpc in size, a result which is substantially different to other studies.
Other observations of the \ciii\ line have been made with IFUs \citep{1989A&A...208L..15A,1994A&A...282...11F} and with narrow-band images \citep{1992ApJ...395L..65R}, but M98 are the only authors claiming to have seen an arc.

This paper aims to resolve the discrepancy between the size of the BELR implied by M98 and current QSO BELR sizes deduced from observations. If BELRs are the size reverberation mapping suggests, then there will be no arc of lensed emission because the size of the macro-lens caustic, and position of the source within the caustic, are very well known \citep[e.g.][]{2002MNRAS.334..621T}. Instead, the BELR will be microlensed.

In section \ref{sec:obsdat}, the observations of Q2237+0305 using the Gemini GMOS IFU are described, along with the data reduction tasks and details of the image extraction process.
In section \ref{sec:psf_sub} we create a PSF from the data and use it to subtract the QSO images, showing that the QSO images of the BELR are unresolved.
In section \ref{sec:flux_ratios} the flux ratios of the BELR images and the surrounding continua are calculated.
We compare to previous work and conclude that the BELR must be microlensed.
In section \ref{sec:sizeBELR} an estimate of the size of the BELR is made, and an upper limit to the size of the continuum source is set. Finally, we summarise in section \ref{sec:summary} and discuss some future prospects.

Throughout this paper, a cosmology with $H_0=70$ km s$^{-1}$ Mpc$^{-1}$, $\Omega_m = 0.3$ and $\Omega_{\Lambda}=0.7$ is used.

\begin{figure}
\begin{centering}
\includegraphics[scale=0.8]{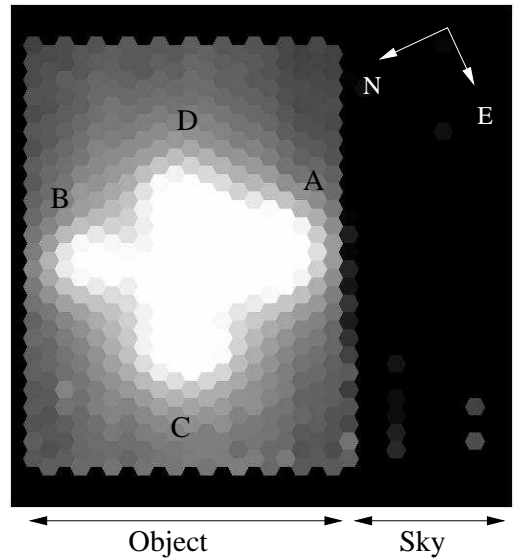}
\caption{Total flux in each lenslet for one of the exposures (the contrast is not linear).
Each image of the QSO is well resolved and separated by at least 5 lenslets from one another. The left section of the image is the sky array, which images a section of the sky approximately 1 arcmin away from the object array. Image labels follow \citet{1988AJ.....95.1331Y}.
}
\label{fig:ifu_lenslets}
\end{centering}
\end{figure}

\section{Observations and data reduction}
\label{sec:obsdat}

2237+0305 was observed on 2002 July 16 and 17 (program ID: GN-2002A-Q-40) using the GMOS IFU on the 8m Gemini North telescope. Three 30min exposures were taken on the 16th and five 30min exposures on the 17th using a random dither pattern with average step size $\sim0.1$ arcsec.
The R400 grating was used which gave a useful spectral range between 450-800 nm with resolution 0.068 nm/pixel.

The GMOS IFU \citep{2002PASP..114..892A} is an array of hexagonal lenslets with projected diameter 0.2 arcsec.
We used the IFU in `one-slit' mode which images a $3.5 \times 5$ arcsec section of the sky.
An additional $5 \times 1.25$ arcsec array, separated by approx 1 arcmin from the main array, is used for sky subtraction.

The five exposures taken on 2002 July 17 are of much better quality than the three from the previous night, so we use only those in the following analysis. The seeing during these observations was 0.65 arcsec. Fig. \ref{fig:ifu_lenslets} shows an image of the total flux in each lenslet for one of the exposures.
 
The data were reduced with the Gemini \textsc{iraf} package v1.4 using \textsc{iraf} v2.11.3b.
The reduction tasks include bias subtraction, flatfielding, arc calibration, spectra extraction, sky subtraction and flux calibration.
Cosmic ray rejection was performed using custom \textsc{iraf} scripts based on simple median filtering.
This produced cleaner cosmic ray rejection than the Gemini \textsc{iraf} task `gscrrej'.
The images were extracted independently and 5 data cubes were generated using `gfcube' which subsamples the hexagonal array into a 0.1 arcsec pixel$^{-1}$ rectangular array.
 
Fig. \ref{fig:ifu_lenslets} shows that the QSO images are well separated and there is good sampling (at least 5 lenslets) between each QSO image and about 5 lenslets between the galaxy centre and the nearest QSO image (image C).

\begin{figure}
\includegraphics[scale=0.42]{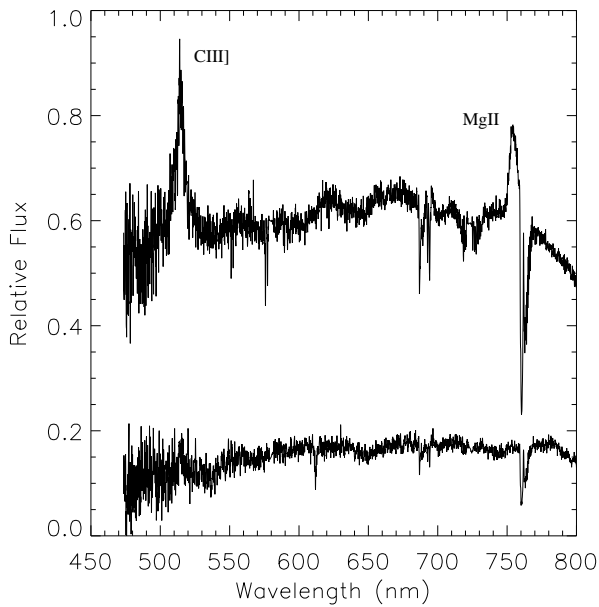}
\caption{Example spectra from an exposure. The top spectrum is from the brightest region of QSO image A, while the bottom spectrum is from the centre of the lens galaxy.}
\label{fig:example_spectra}
\end{figure}

Fig. \ref{fig:example_spectra} shows example spectra taken from the centre of QSO image A and from the centre of the lens galaxy.
The redshifted \ciii\ and \mgii\ lines are clearly seen in the QSO spectrum but are barely visible in the spectrum from the galaxy. The reddest edge of the \mgii\ broad line is absorbed by the atmosphere around 760nm. These spectra show that the QSO images and galaxy have been well resolved by the IFU. We note that the instrument was not corrected for atmospheric dispersion, which shifted the locations of the QSO images by approximately one lenslet at the wavelengths of the two emission lines.
However, the following analysis is not affected by the dispersion. 
\begin{figure*}
\includegraphics[scale=0.6]{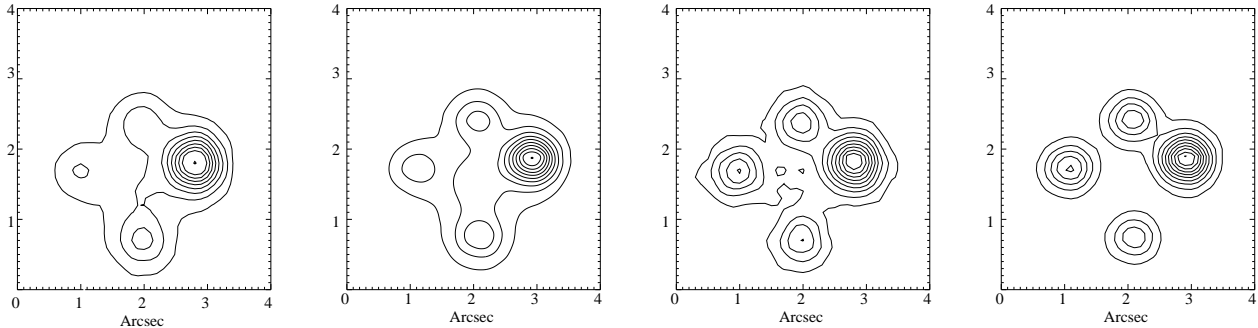}
\caption{Contours of the combined images. Left to right: \ciii\ continuum, \mgii\ continuum, \ciii\ line and \mgii\ line. The line images have had the continuum subtracted as described in the text. Contours are spaced linearly in intervals of 10\% of the peak in each image.}
\label{fig:lineflux_contours}
\end{figure*}

For each exposure, the \ciii\ and \mgii\ line flux were summed between (507.9nm, 520.8nm) for \ciii\ and (747.8nm, 758.5nm) for \mgii. We also calculated the total continuum flux for \ciii\ between (501.3nm, 507.8nm and 520.9nm, 527.3nm) and \mgii\ between (742.2nm, 747.8nm and 768.4nm, 774.0nm avoiding the sky absorption) resulting in four images per exposure: one each for the \ciii\ line--plus--continuum, \mgii\ line--plus--continuum, the continuum around the \ciii\ line and the continuum around the \mgii\ line.
We then subtracted the continuum images from the line--plus--continuum images to make images of the integrated line flux.
We use the integrated flux in the line rather than the equivalent width because an equivalent width measurement is susceptible to different lensing magnification between the BELR and the continuum regions in the QSO.

The continuum subtraction process was very clean, leaving no trace of the galaxy emission in the centre of the images.
In each line and continuum image, the four QSO images are clearly resolved with a stable PSF.
There were some small variations in the PSF between images, but it was reassuring to find that each of the line flux images looked almost identical except for the dither step offsets.
The \ciii\ images were somewhat noisier than the \mgii\ due to the lower signal/noise in the blue end of the spectra.
We then combined the total line flux images and the continuum images, using the known dither steps, for all five \ciii\ and \mgii\ images, resulting in two final line flux images (hereafter `\ciii\ line' and `\mgii\ line') and two final continuum images (hereafter `\ciii\ cont' and `\mgii\ cont').
The resulting continuum and line images are shown in Fig. \ref{fig:lineflux_contours}.

It can be seen immediately that the QSO images are well separated and that the galaxy has disappeared in the line images.
The arc joining A and D is consistent with PSF overlap, not a lensed arc which we would expect to be much more circular.
There is certainly nothing resembling the relatively bright arc as seen by M98.
 
In the next section, we subtract the QSO images to look for any faint resolved line emission.
 
\section{PSF subtraction}
\label{sec:psf_sub}
Due to varying atmospheric conditions, there are likely to be large differences between the standard star used for flux calibration and the PSF of the science observations. We therefore elected to reconstruct the PSF from the four QSO images. The light from the lensing galaxy is already subtracted in the line images, so extraction of the PSF is a matter of deconvolving four point sources. We achieve this through an iterative approach and produced a PSF for each of the line images.

\begin{figure}
\includegraphics[scale=0.5]{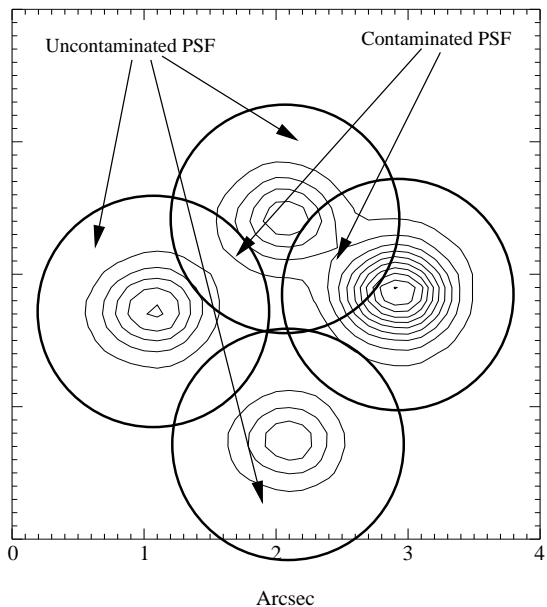}
\caption{Apertures used over each QSO image in construction of the PSF. The final PSF was a combination of the four areas of uncontaminated PSF.}
\label{fig:psf_areas}
\end{figure}

\begin{figure}
\hspace{5mm}
\includegraphics[scale=0.25]{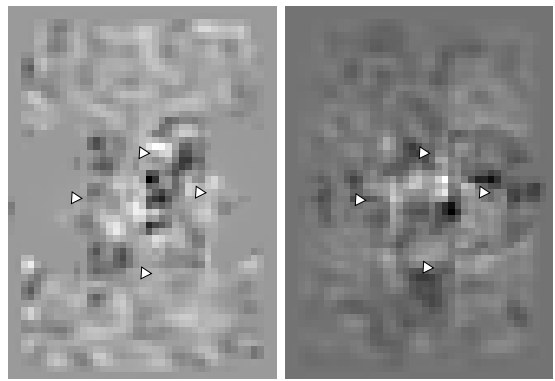}
\caption{Residuals of the \ciii\ (left) and \mgii\ (right) PSF subtraction. The white triangles show the positions of the QSO images before subtraction.}
\label{fig:psf_sub_resid}
\end{figure}

First, we constructed a trial PSF by defining apertures of 0.8-0.9 arcsec (depending on brightness) for each image, as shown in Fig. \ref{fig:psf_areas}. The regions covered by only a single aperture are considered 'uncontaminated', and are shifted and combined to produce our trial PSF, with each segment normalised according to the relative fluxes in the regions where the sampled regions overlap.
This PSF was then used to find the fluxes of the four images using the $\chi^2$ statistic to minimise residuals.
These fluxes were used to create four separate subtracted images. In each of these a single QSO image is left unsubtracted. We then repeated the first step, constructing a new PSF using the same 'uncontaminated' regions to sample each of these four lens images.
The trial PSF converged after only a few iterations, resulting in the residual maps shown in Fig. \ref{fig:psf_sub_resid}. Note that any flux from a possible faint arc between images A and D fall into an overlapping region, so does not affect the result.

Using the constructed PSF, a scaled PSF was subtracted from each QSO line image such that the $\chi^2$ residual over the image was minimised.
The QSO-subtracted residual for each line image is shown in Fig. \ref{fig:psf_sub_resid}.
(We show the residuals as images because contour plots are difficult to interpret.)
The peak(RMS) of the residual, compared to the peak in the line image, was 9.9(1.4) and 4.5(0.5) percent for \ciii\ and \mgii\ respectively. This compares well to the RMS noise in the line images (computed using pixel values in the corners of the images far from the bright regions), of 1.2 and 0.2 percent.
We note that the residuals were worst between the QSO images (i.e. around the centre of the lens galaxy) where the most PSF overlap occurs. In the \ciii\ residual, the region between the images is slightly over-subtracted, whereas in the \mgii\ image it is slightly under-subtracted. These are artifacts of an imperfect PSF subtraction. Most importantly, there is no evidence for any extended emission (arcs or rings) joining the QSO images. We conclude, therefore, that the images of the QSO BELR are unresolved.

The difference between this data and M98 is very important for the determination of size of the BELR.
Possible explanations for why an arc was seen by M98 include poorer spatial resolution, seeing or telescope/instrumental effects.
The INTEGRAL IFU \citep{1998fopa.proc..149A} (with SB1 package) used by M98 has circular lenslets separated by 0.5 arcsec and a 65\% filling factor.
Since flux is lost between fibres and the separation between images A/D and B/D is $\le 1.2$ arcsec, we would argue that the total flux and positions of the QSO images cannot be determined reliably in 0.7\arcsec\ seeing because the PSF is under-sampled. There is (in the best case) only one fibre between images A/D and B/D with INTEGRAL, so flux from surrounding images must leak into this fibre in 0.7 arcsec  seeing.
After interpolating, such leakage could easily look like an arc.
The GMOS IFU, on the other hand, has at least five lenslets between images and our data have not been interpolated after resampling the 0.2 arcsec hexagonal lenslets into a 0.1 arcsec pixel$^{-1}$ rectangular array.

\section{Flux ratios}
\label{sec:flux_ratios}
In the previous section the BELR was found to be unresolved in both the \ciii\ and \mgii\ lines for the QSO.
A test for any significant difference in the size and/or location of emitting regions for the two lines is in the measurement of the flux ratios between images for each line.
If the emitting regions are of significantly different size (or not co-located on the line-of-sight), then microlensing will produce different magnifications.
Microlensing has already been shown to change the equivalent width of the emission lines over time \citep{1998MNRAS.295..573L}, however the equivalent width can be changed by microlensing of the continuum region alone.

\begin{figure}
\vspace{-5mm}
\includegraphics[scale=0.4]{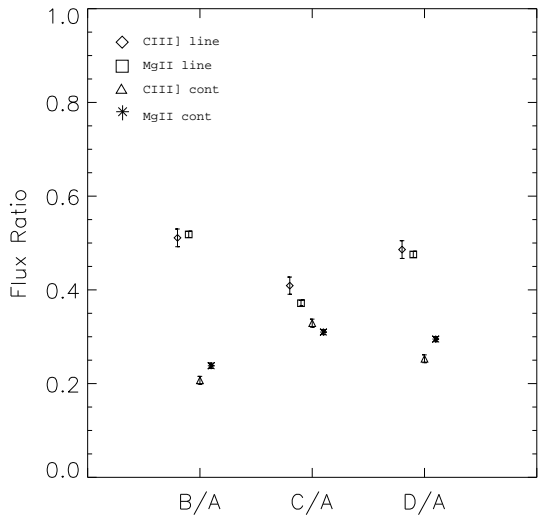}
\caption{Flux ratios for the line and continuum images, not corrected for extinction.}
\vspace{-5mm}
\label{fig:flux_ratio_noext}
\end{figure}

Table \ref{tab:photometry} shows the relative amplitude of the PSF used to subtract each QSO image in section \ref{sec:psf_sub}. Given the QSO images are unresolved, the PSF magnitude is a photometric measurement which can then be used to calculate the ratio of fluxes (hence magnifications) between the QSO images.
To calculate the flux ratios of the continuum data, the same procedure used for the PSF subtraction in section \ref{sec:psf_sub} was employed, with an additional component for the galaxy bulge light. We used a de Vaucouleurs model with scale length \mbox{4.1 \arcsec}, position angle 77\degr and axis ratio 0.69 \citep{1988AJ.....95.1331Y,1996MsT..........2S} which is convolved with the constructed PSF. All five components are fit to minimise the $\chi^2$ residual as before. The continuum flux measurements are also shown in Table \ref{tab:photometry}. The (not de-reddened) flux ratios for these measurements are plotted in Fig. \ref{fig:flux_ratio_noext}.

\begin{table*}
\begin{tabular}{l|llll}
Region of spectrum & A & B & C & D \\ \hline
\ciii\ line & $\equiv 1 \pm 1.7\%$ &$0.511 \pm 3.3\%$ & $0.409 \pm 4.2\%$ & $0.486 \pm 3.5\%$  \\
\mgii\ line & $\equiv 1 \pm 0.6\%$ &$0.518 \pm 1.1\%$ & $0.371 \pm 1.5\%$ & $0.476 \pm 1.2\%$  \\
\ciii\ cont & $\equiv 1 \pm 0.82\%$ & $0.207 \pm 4.0\%$ & $0.329 \pm 2.5\%$ & $0.253 \pm 3.2\%$ \\
\mgii\ cont & $\equiv 1 \pm 0.55\%$ &$0.238 \pm 2.3\%$ & $0.310 \pm 1.7\%$ & $0.295 \pm 1.8\%$
\end{tabular}
\caption{Photometry (not extinction corrected) for the QSO images in both emission line strength and continuum surrounding the emission lines. The measurements are normalised to image A. Errors are $1\sigma$.}
\label{tab:photometry}
\end{table*}

We corrected for extinction using the scheme of \citet{2000ApJ...545..657A}.
We used their published $A_V$ values for the \ciii\ line and continuum flux measurements and calculated the extinction $A$(750nm) using the R=3.1 extinction curve from \citet{1999PASP..111...63F} for the \mgii\ measurements.
\citet{1999ApJ...523..617F} found a significantly higher value for R (5.29), but it makes little difference to the results at the wavelengths of interest.

Table \ref{tab:flux_ratios_corrected} shows the extinction corrected flux ratios along with the mid-IR, radio flux, \ciii\ narrow band and $H_{\beta}$ ratios from \citet{2000ApJ...545..657A,1996AJ....112..897F,1992ApJ...395L..65R,2004ApJ...607...43M} respectively. The radio and mid-IR fluxes are consistent with each other and are generally thought to represent the magnification ratios free of any microlensing. The \ciii\ narrow band images, used by \cite{1992ApJ...395L..65R} as a baseline for detecting microlensing, are inconsistent with the radio/mid-IR even though the B/A flux alone is consistent.
Since the \ciii\ narrow band flux ratios have changed, they must be influenced by microlensing.
Surprisingly, the $H_{\beta}$ broad line is not really consistent with any other observation although the uncertainty due to extinction means the difference has low statistical significance.
We note that the dominant component of the error is due to the uncertainty in the extinction. Nevertheless, it is clear that the B/A and D/A flux ratios are different between the line and continuum images and other measurements.
Furthermore, the flux ratios of the \ciii\ and \mgii\ line images themselves are consistent within errors.
We also argue that the flux ratios cannot be due to gravitational millilensing by matter between the QSO and lens. For millilensing to produce different magnifications between the continuum and BELR regions, the BELR must have angular size of order milli-arcseconds, which is $\sim 10$pc in the source plane. This is too large to be consistent with reverberation mapping results. However, it is the change in the B/A flux ratio of the \ciii\ line between the data presented here and those of \citet{1992ApJ...395L..65R} that most convincingly rules out millilensing. The \emph{raw} (not de-reddened) B/A flux ratio has changed from $1.04 \pm 0.04$ to $0.51 \pm 0.02$ during the 11 years between observations. Regardless of the actual amount of extinction for each image, the extinction will remain constant over this time\footnote{A hypothetical dust cloud would need to be travelling at least $\sim 1000$km s$^{-1}$ to traverse an image in 11 years and would produce a noticeable long-term effect.}. Hence, microlensing is the only process which can change the B/A flux ratios so substantially in such a short time.

\begin{table}
\begin{tabular}{l|lll}
Region & B/A & C/A & D/A \\ \hline
\ciii\ line & $0.492 \pm 25\% $ & $0.602 \pm 31\% $ & $0.624 \pm 28\% $ \\
\mgii\ line  & $0.505 \pm 17\% $ & $0.487 \pm 22\% $ & $0.566 \pm 20\% $ \\
\ciii\ cont & $0.200 \pm 18\% $ & $0.484 \pm 31\% $ & $0.325 \pm 28\% $ \\
\mgii\ cont  & $0.232 \pm 18\% $ & $0.406 \pm 22\% $ & $0.351 \pm 20\% $ \\ \hline
Mid IR     & $1.11 \pm 10\%  $ & $0.59 \pm 15\% $  & $1.00 \pm 10\% $ \\
8 GHz      & $1.08 \pm 0.27 $  & $0.55 \pm 0.21 $  & $0.77 \pm 0.23 $ \\
\ciii\ NB & $1.04 \pm 0.04$ & $0.63 \pm 0.03$ & $0.55 \pm 0.03$ \\
$H_{\beta}$ line & $0.376 \pm 0.007$ & $0.387 \pm 0.007 $ & $0.461 \pm 0.004$
\end{tabular}
\caption{Top: flux ratios \citep[extinction corrected using][]{2000ApJ...545..657A} between line flux and continuum images. Errors are $1\sigma$. Bottom: flux ratios from previous work in the mid-IR \citep{2000ApJ...545..657A}, radio \citep{1996AJ....112..897F}, \ciii\ with narrow-band filters \citep{1992ApJ...395L..65R} and near-IR \citep{2004ApJ...607...43M}. }
\label{tab:flux_ratios_corrected}
\end{table}

This leads to the following conclusions: i) the sizes and locations of the BELRs for the \ciii\ and \mgii\ lines are very similar, because they appear to be subject to the same amount of microlensing, ii) the BELR must be microlensed because the flux ratios for the line images are substantially different to the radio, mid-IR and previous narrow band fluxes and iii) the continuum region is a substantially different size to the BELR.

It is interesting that the line flux ratios of the $H_{\beta}$ broad line in \citet{2004ApJ...607...43M} differ from this work (albeit at a low significance). The two independent sets of observations were separated in time by only 25 days which should be too short for significant changes in either the luminosity or the magnification of the BELR to occur.
This suggests that the section of the BELR emitting $H_{\beta}$ may be different to that of \ciii\ and \mgii, consistent with a stratified BELR model.
It is possible that the results in \citet{2004ApJ...607...43M} were affected by the poor PSF of those observations. Nevertheless, it is intriguing and certainly warrants further investigation.

\section{Limits on the size of the BELR and continuum region}
\label{sec:sizeBELR}
For a large source (relative to the projected Einstein radius of the microlensing stars), the flux ratio should be that predicted by the macro-model for the lensing galaxy (approx 0.7-1.0 depending on the model. [E.g: \citet{2002MNRAS.334..621T} for a surface brightness based model, and \citet{1994AJ....108.1156W} for a parametric model].
As the source size decreases, the impact of microlensing on the flux ratio increases.

An estimate can be made of the sizes of the BELR and the continuum region using fig. 2 from \citet{2002MNRAS.331.1041W} which plots the cumulative probability of observing a B/A flux ratio as a function of the source size (measured in units of the projected microlens Einstein radius $\eta_0 = \sqrt{\frac{D_{ds} D_s 4 G M}{D_d c^2}}$, which is $0.06 h_{70}^{1/2} \left ( M/M_{\odot} \right )^{1/2} $pc in this case).
This figure shows that for a source size of 0.4$\eta_0$, the flux ratio should be $\le 0.5$ 80 percent of the time and $\le 0.25$ 20 percent of the time. Likewise, if the source size is 1.6$\eta_0$, then the flux ratio should be $\le 0.75$ 80 percent of the time and $\le 0.4$ 20 percent of the time. If it is 6.4$\eta_0$ or larger then the flux ratio should never be $\le 0.5$.
Hence, there is only a limited range of values for the source size that are likely to produce a B/A flux ratio of 0.5, and an upper limit to the size which can produce a ratio of 0.25 or 0.5.
A cumulative probability of 0.5 gives the most likely value for the flux ratio which, interpolating between the S=0.4 and S=1.6 curves, is S $\sim 1 \eta_0$ for a flux ratio of 0.5.

Reading from this figure, the following conclusions can be made:
i) for the B/A flux ratio to be 0.5, the emitting region is between 0.4 and 6.4 $\eta_0$ with 80 percent confidence and is likely to be $\sim 1 \eta_0 = 0.06$pc, 
and ii)
for the B/A flux ratio to be $0.25$, the continuum region is $\le 0.4 \eta_0 \le 0.02 $ pc with 80 percent confidence and is likely to be $\ll 0.4 \eta_0$.

These determinations for the size of the BELR are inconsistent with M98.
If the BELR has a physical size of order kpc as suggested by M98, then the flux ratios cannot be affected by microlensing or millilensing and the lensed images of the QSO BELRs should be distorted by the macro lensing.
Hence we conclude that the BELR cannot be as large as M98 suggest.
We note that the estimates presented here are based on the assumptions in \citet{2002MNRAS.331.1041W}, in particular that the surface brightness of the emitting region is a smooth Gaussian.

\begin{figure}
\vspace{-5mm}
\includegraphics[scale=0.4]{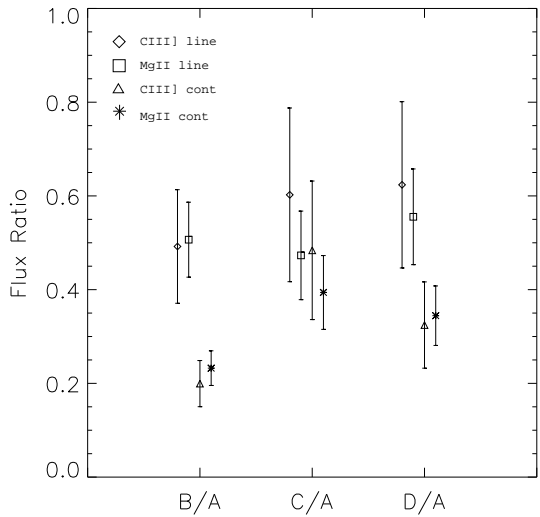}
\vspace{-5mm}
\caption{Flux ratios for the line and continuum images after extinction correction.}
\label{fig:flux_ratio_extcorr}
\end{figure}

\section{Summary and discussion}
\label{sec:summary}

We observed the Q2237+0305 using the GMOS IFU on the 8m Gemini North telescope.
The GMOS IFU is an ideal instrument for observing lensed QSOs with hexagonal lenslets separated by 0.2 arcsec.
Images were made of the integrated \ciii\ and \mgii\ broad line emission and the surrounding continua.
After constructing a PSF from the (continuum subtracted) line images, the QSO images were subtracted, leaving no significant residual (resolved) flux.
We concluded that the arc of emission seen by \citet{1998ApJ...503L..27M} was probably an artifact generated by poor spatial sampling in the INTEGRAL IFU.

The flux ratios of the lensed QSO images were calculated for the both the broad line and the continuum images including a correction for extinction.
The flux ratios for the broad line images are consistent with each other, but are inconsistent with both the continuum flux ratios (from this data) and previously published work.
This strongly suggests that the BELR must be microlensed, but to a different degree than the QSO continuum region.
The identical flux ratios of the \ciii\ and \mgii\ lines also suggest the emitting regions for the two lines are the same size and located along the same line-of-sight.
Using previous results on the likelihood of observing a particular B/A flux ratio, we estimated the size of the \ciii /\mgii\ BELR to be $\sim 0.06 h_{70}^{1/2}$ pc in size and the continuum region is $\le 0.02 h_{70}^{1/2}$ pc.
The continuum measurement is consistent with \citet{2000MNRAS.318..762W} who estimate the size of the continuum region to be $\le 0.01$pc based on the properties of the image light-curves, not differential magnification.

This measurement of the BELR size is broadly consistent with reverberation mapping results for high-luminosity AGNs. The result could be strengthened further by continued monitoring of the broad lines.
 Our results indicate that this is an area of study which warrants more attention and there are several other lensed QSOs which could benefit from a study such as this. Using QSOs with a range of redshifts (hence projected microlens Einstein radii) and luminosities will provide further constraints on the BELR by probing different size scales at lower optical depths.

\section*{Acknowledgements}
RBW thanks Kathy Roth, the Gemini contact scientist for GN-2002A-Q-40, for support and helpful suggestions during the phase II process. RBW also thanks Cathryn Trott, Geraint Lewis and Stuart Wyithe for suggestions and help during preparation of this paper.
RBW is grateful for support from the David Hay Memorial Fund.
\newcommand{\apj}{ApJ}
\newcommand{\nat}{Nat}
\newcommand{\mnras}{MNRAS}
\newcommand{\aj}{AJ}
\newcommand{\pasp}{PASP}
\newcommand{\aap}{A\&A}
\newcommand{\apjl}{ApJ}
\newcommand{\araa}{ARA\&A}
\newcommand{\pasj}{PASJ}

\label{lastpage}

\end{document}